\documentclass[10pt]{elsarticle}

\usepackage{bm}
\usepackage{mathtools}
\usepackage{graphicx}
\usepackage{multirow}
\usepackage{amsthm,nccmath}

\usepackage{algorithm}
\usepackage{algpseudocode}
\usepackage{algpascal}
\usepackage{textcomp}

\usepackage{tikz}

\usepackage{smartdiagram}
\usepackage[utf8]{inputenc}
\usetikzlibrary{arrows,positioning}
\usepackage{makecell} 
\setcellgapes{5pt}
\DeclarePairedDelimiter{\ceil}{\lceil}{\rceil}

\usepackage{amsmath,amsfonts,amssymb}

\smartdiagramset{border color=none,
	set color list={blue!50!cyan,green!60!lime,orange!50!red,red!80!black},
	back arrow disabled=true, module minimum width = 4}

\smartdiagramset{module minimum height=1cm,
	module minimum width=0.8cm,
	module x sep=2}

\usepackage{hyperref}

\journal{}

\begin{document}
	
	\begin{frontmatter}
		
		\title{New bootstrap tests for categorical time series. A comparative study}


		\author[mymainaddress]{\'Angel L\'opez-Oriona\corref{mycorrespondingauthor} (ORCID 0000-0003-1456-7342)}
		\ead{oriona38@hotmail.com, a.oriona@udc.es}
		
		\author[mymainaddress]{Jos\'e A. Vilar (ORCID 0000-0001-5494-171X)}
		\cortext[mycorrespondingauthor]{Corresponding author}
		\ead{jose.vilarf@udc.es}
		
		\author[dursoaddress]{Pierpaolo D'Urso (ORCID 0000-0002-7406-6411)}
		\ead{pierpaolo.durso@uniroma1.it}
		
		\address[mymainaddress]{Research Group MODES, Research Center for Information and Communication Technologies (CITIC), University of A Coru\~na, 15071 A Coru\~na, Spain.}
	   \address[dursoaddress]{Department of Social Sciences and Economics, Sapienza University of Rome, P. le Aldo Moro 5, Roma, Italy.}

		
		\begin{abstract}
			The problem of testing the equality of the generating processes of two categorical time series is addressed in this work. To this aim, we propose three tests relying on a dissimilarity measure between categorical processes. Particular versions of these tests are constructed by considering three specific distances evaluating discrepancy between the marginal distributions and the serial dependence patterns of both processes. Proper estimates of these dissimilarities are an essential element of the constructed tests, which are based on the bootstrap. Specifically, a parametric bootstrap method assuming the true generating models and extensions of the moving blocks bootstrap and the stationary bootstrap are considered. The approaches are assessed in a broad simulation study including several types of categorical models with different degrees of complexity. Advantages and disadvantages of each one of the methods are properly discussed according to their behavior under the null and the alternative hypothesis. The impact that some important input parameters have on the results of the tests is also analyzed. An application involving biological sequences highlights the usefulness of the proposed techniques. 
		\end{abstract}
		
		
	\end{frontmatter}
	
	
	\section{Introduction}\label{sectionintroduction}
	
	The problem of comparing two time series arises in a natural way in multiple fields, including artificial intelligence, economics, computer science, biology, medicine or chemistry, among others. For instance, an investor often has to determine if two particular assets show the same behavior over time based on historical data. In medicine, it is usually interesting to find out to what extent ECG signals from different subjects exhibit similar patterns. A broad variety of data mining and statistical methods have been proposed to address this kind of problems, including clustering \cite{aghabozorgi2015time}, classification \cite{abanda2019review}, outlier detection \cite{blazquez2021review}, and comparisons through hypothesis tests \cite{lopez2022bootstrap}. It is worth highlighting that these approaches have mainly focused on real-valued time series \cite{maharaj1999comparison, d2009autocorrelation, lafuente2016clustering, lopez2021outlier, lopez2021f4}, while the study of time series with alternative ranges, for instance, categorical time series (CTS), has received much less attention \cite{fruhwirth2010model, lopez2023hard}. This is surprising, since CTS are frequently used for important tasks. Some illustrative examples are the stochastic modeling of DNA sequence data \cite{fokianos2003regression, weiss2008measuring}, the analysis of EEG sleep state scores \cite{stoffer2000spectral}, and the use of hidden Markov models (HMM) to analyze protein sequences \cite{krogh1994hidden}. 
	
	Frequently, these techniques require to evaluate dissimilarity between time series, which is not a simple task due to the dynamic character of these objects. In fact, the problem of determining a proper distance measure between time series has become an important research topic. In the real-valued setting, \cite{lafuente2016clustering} provided a clustering algorithm for time series based on an innovative distance comparing the so-called quantile autocovariance functions. Other dissimilarity criteria recently proposed to construct clustering procedures include distances between: estimated GARCH coefficients \cite{d2016garch}, B-splines representations \cite{d2021robust} and estimated conditional moments \cite{cerqueti2021model}, among many others. Several methods employing distance measures have also been proposed for classifying real-valued series \cite{mei2015learning, lucas2019proximity}. The definition of a suitable dissimilarity becomes even more complex in the categorical context, since most of the standard tools used to deal with real-valued time series (e.g., the autocorrelation function) are no longer valid when analyzing CTS. \cite{garcia2015framework} introduced a dissimilarity between CTS which evaluates both closeness between raw categorical values and proximity between dynamic patterns. \cite{lopez2023hard} proposed two novel feature-based distances between categorical series measuring discrepancy between their marginal distributions and their underlying serial dependence patterns. In both works, the corresponding metrics are applied to perform CTS clustering. 
	
	The aim of the present work is to introduce procedures to test that two categorical processes are equal in terms of marginal distributions and serial dependence structures. Specifically, let $\{X_t^{(1)}, t \in \mathbb{Z}\}$ and $\{X_t^{(2)}, t \in \mathbb{Z}\}$ be two independent stationary categorical processes with range $\mathcal{V}=\{1,\ldots,r\}$ and denote by $\boldsymbol \pi^{(1)}=(\pi_1^{(1)}, \ldots, \pi_r^{(1)})$ and $\boldsymbol \pi^{(2)}=(\pi_1^{(2)}, \ldots, \pi_r^{(2)})$, respectively, the corresponding vectors of marginal probabilities, that is, $\pi_h^{(i)}=P\big(X_t^{(i)}=h\big)$, $i=1,2$, $h=1,\ldots,r$. In addition, given a lag $l \in \mathbb{Z}$ and $j,k \in \mathcal{V}$, let $p_{jk}^{(i)}(l)$ be the corresponding lagged joint probability for process $\{X_t^{(i)}, t \in \mathbb{Z}\}$, that is, $p_{jk}^{(i)}(l)=P(X_t^{(i)}=j, X_{t-l}^{(i)}=k)$, $i=1,2$. The null hypothesis we consider can be stated as
	
	\begin{equation}\label{hypothesis1}
		H_0: \boldsymbol \pi^{(1)}=\boldsymbol \pi^{(2)} \, \, \, \, \text{and} \, \, \, \, p_{jk}^{(1)}(l)=p_{jk}^{(2)}(l) \, \, \, \, \forall (j, k, l) \in \mathcal{V}^2 \times \mathbb{Z}. 
	\end{equation} 

In order to perform the hypothesis test in \eqref{hypothesis1}, we consider three distance measures between categorical stochastic processes, whose estimates were employed by \cite{lopez2023hard} to perform clustering of CTS. Two of these dissimilarities are based on extracted features describing the marginal properties and the serial dependence structures of both stochastic processes. The remaining metric relies on the coefficients defining a given categorical model. In the first two cases, the distances take the value of 0 when the null hypothesis is true, which makes the estimates of these metrics a reasonable tool to carry out the test in \eqref{hypothesis1}. It is worth highlighting that the computation of the asymptotic distribution of these estimates under the null hypothesis is a very challenging problem if a specific generating structure is not assumed, so resampling techniques can be considered to perform the test.

Based on previous considerations, three bootstrap methods are proposed in this work to approximate the null distribution of the considered estimates. The first technique is a parametric test which assumes a specific class of categorical model for both stochastic processes. The crucial step of this procedure is based on the generation of time series from a process which contains information about both original series in equal measure. The remaining tests are extensions of two bootstrap approaches specifically designed to deal with dependent data, namely the moving blocks bootstrap (MBB) \cite{kunsch1989jackknife, liu1992moving} and the stationary bootstrap (SB) \cite{politis1994stationary}. In both cases, the key
principle is to generate pseudo-series with the aim of mimicking the distribution under the null hypothesis of the corresponding estimates without assuming specific parametric models for the generating processes. The bootstrap approaches based on the three metrics are compared in terms of size and power by means of a broad simulation study. Several types of generating processes are considered under the null and alternative hypotheses. Finally, an interesting application involving biological sequences highlights the usefulness of the proposed methods. It is worth remarking that, although there exist many statistical tests for assessing dissimilarity between the generating processes of two time series \cite{swanepoel1986comparision, maharaj2002comparison, mahmoudi2017testing}, most of them focus on the real-valued setting. In fact, to the best of our knowledge, there exist no works in the literature dealing with the comparison of the generating structures of two categorical series.

The rest of the article is organized as follows. The three considered distances between categorical processes are defined in Section \ref{sectionbackground} after introducing some features measuring serial dependence within these type of processes. The three bootstrap techniques to carry out the test in \eqref{hypothesis1} are presented in Section \ref{sectionbootstraptests}. A description of the simulation experiments  performed to compare the proposed tests is provided in Section \ref{sectionsimulationstudy} along with the corresponding results and discussion. Section \ref{sectionapplication} contains the application of the bootstrap tests and Section \ref{sectionconclusions} concludes.

	\section{Background on three dissimilarity measures for categorical series}\label{sectionbackground}
	
	Hereafter, $\{X_t, t \in \mathbb{Z}\}$ (or just $X_t$) denotes a categorical stochastic process taking values on a number $r$ of  unordered qualitative categories, which are coded from 1 to $r$ so that the range of the process can be seen as $\mathcal{V}=\{1,\ldots,r\}$. It is assumed that $X_t$ is bivariate stationary, that is, the pairwise joint distribution of $(X_{t}, X_{t-l})$ is invariant in $t$ for arbitrary $l$ (see \cite{weiss2008measuring}). The marginal distribution of $X_t$ is denoted by $\bm{\pi}=(\pi_1, \ldots, \pi_r)$, with $\pi_j=P(X_t=j)$, $j=1,\ldots,r$. Fixed $l \in \mathbb{N}$, we use the notation $p_{ij}(l)=P(X_t=i, X_{t-l}=j)$, with $i,j \in \mathcal{V}$, for the lagged joint probability and the notation $p_{i|j}(l)=P(X_t=i|X_{t-l}=j)=p_{ij}(l)/\pi_j$ for the conditional lagged probability. 
	
	Next, we introduce different sets of features that can be used to describe the process $X_t$ and, afterwards, we present the dissimilarity measures based on the corresponding sets of features. 
	
	\subsection{Structural features for categorical processes}\label{subsectionstructuralfeatures}
	
	In order to extract suitable features characterizing the serial dependence of a given categorical process, we first start by defining the concepts of perfect serial independence and dependence for a categorical process. Following \cite{weiss2008measuring}, we have perfect serial independence at lag $l \in \mathbb{N}$ if and only if $p_{ij}(l)=\pi_i\pi_j$ for any $i,j \in \mathcal{V}$. On the other hand, we have perfect serial dependence at lag $l \in \mathbb{N}$ if and only if the conditional distribution $p_{\cdot|j}(l)$ is a one-point distribution for any $j \in \mathcal{V}$. This way, in a perfect serially independent process, knowledge about $X_{t-l}$ does not help at all in predicting the value of $X_{t}$. Conversely, in a perfect serially dependent process, the value of $X_{t}$ is completely determined from $X_{t-l}$. 
	
	There are several association measures that describe the serial dependence structure of a categorical process at lag $l$. One of such measures is the so-called Cramer\textquotesingle s $v$, which is defined as 
	\begin{equation}\label{cramerv}
		v(l)=\sqrt{\frac{1}{r-1}\sum_{i,j=1}^r\frac{(p_{ij}(l)-\pi_i\pi_j)^2}{\pi_i\pi_j}}.
	\end{equation}

The quantity $v(l)$ has range $[0,1]$, with the values 0 and 1 associated with the cases of perfect serial independence and perfect serial dependence at lag $l$, respectively. Note that the numerator appearing in the summation of \eqref{cramerv} measures the deviation of $p_{ij}(l)$ from the case of serial independence between $i$ and $j$ at lag $l$. 
	
	Cramer\textquotesingle s $v$ summarizes the serial dependence levels of a categorical process for every pair $(i,j)$ and $l \in \mathbb{N}$. However, this quantity is not appropriate for characterizing a given stochastic process, since different processes can exhibit the same value of $v(l)$. A better way to characterize the process $X_t$ is by considering the matrix $\bm V(l)=\big(V_{ij}(l)\big)_{1\leq i, j \leq r}$, where 
	\begin{equation}\label{matrixv}
		V_{ij}(l)=	\frac{(p_{ij}(l)-\pi_i\pi_j)^2}{\pi_i\pi_j}.
	\end{equation}
	
	In this way, the $r^2$ elements in the summation of \eqref{cramerv} are separately considered, and a much richer picture of the underlying dependence structure of $X_t$ is available. 
	
	The elements of the matrix $\bm V(l)$ give information about the so-called \textit{unsigned} dependence of the process. However, it is often useful to know whether a process tends to stay in the state it has reached or, on the contrary, the repetition of the same state after $l$ steps is infrequent. This motivates the concept of \textit{signed} dependence, which arises as an analogy of the autocorrelation function of a real-valued process, since such quantity can take either positive or negative values. The reader is referred to \cite{weiss2008measuring, lopez2023hard} for more details about the concepts of unsigned and signed serial dependence.  
	
	
	Since $\bm V(l)$ does not shed light on the signed dependence patterns, it would be valuable to complement the information contained in this matrix by adding features describing signed dependence. In this regard, a common measure of signed serial dependence at lag $l$ is the Cohen\textquotesingle s $\kappa$, which takes the form
	\begin{equation}
		\kappa(l)=\frac{\sum_{j=1}^r(p_{jj}(l)-\pi_j^2)}{1-\sum_{j=1}^{r}\pi_j^2}.
	\end{equation}
	
	Proceeding as with $v(l)$, the quantity $\kappa(l)$ can be decomposed in order to obtain a more detailed representation of the signed dependence pattern of the process. In this way, we consider the vector $\bm{\mathcal{K}}(l)=\big(\mathcal{K}_1(l), \ldots, \mathcal{K}_r(l)\big)$, where each $\mathcal{K}_i(l)$, 	for $i=1,\ldots,r,$ is defined as
	\begin{equation}\label{vectorkappa}
		\mathcal{K}_i(l)=\frac{p_{ii}(l)-\pi_i^2}{1-\sum_{j=1}^{r}\pi_j^2}.
	\end{equation}
	
	In practice, the matrix $\bm V(l)$ and the vector $\bm{\mathcal{K}}(l)$ must be estimated from a $T$-length realization of the process, denoted by $(x_1,\ldots, x_T)$. To this aim, we consider estimators of $\pi_i$ and $p_{ij}(l)$, denoted by $\widehat\pi_i$ and $\widehat p_{ij}(l)$, respectively, defined as  
	\begin{equation}\label{estimates}
		\widehat\pi_i=\frac{N_i}{T} \, \, \, \, \text{and} \, \, \, \,   \widehat p_{ij}(l)=\frac{N_{ij}(l)}{T-l},
	\end{equation}
	where $N_i$ is the number of elements $x_t$ equal to $i$ in the realization $(x_1,\ldots, x_T)$, and $N_{ij}(l)$ is the number of pairs $(x_t,x_{t-l})=(i,j)$ in the realization $(x_1,\ldots, x_T)$. Hence, estimates of $\bm V(l)$ and $\bm{\mathcal{K}}(l)$, denoted by $\widehat{\bm V}(l)$ and $\widehat{\bm{\mathcal{K}}}(l)$, respectively, can be obtained by considering the estimates $\widehat\pi_i$ and $\widehat p_{ij}(l)$ in \eqref{matrixv} and \eqref{vectorkappa}. This leads directly to estimates of $v(l)$ and $\kappa(l)$, denoted by $\widehat v(l)$ and $\widehat \kappa(l)$, respectively, whose asymptotic distributions have been studied for the i.i.d. case by \cite{weiss2013serial} and \cite{weiss2011empirical}, respectively. Note that, by considering $\widehat{\bm V}(l)$ and $\widehat{\bm{\mathcal{K}}}(l)$, a complete picture of the serial dependence patterns of a CTS is provided.

	An alternative way of describing the dependence structure of the process $\{X_t, t \in \mathbb{Z}\}$ is by taking into consideration its equivalent representation as a multivariate binary process. The so-called \textit{binarization} of $\{X_t, t \in \mathbb{Z}\}$ is obtained as follows. Let $\bm e_1, \ldots, \bm e_r \in \{0,1\}^r$ be unit vectors such that $\bm e_k$ has all its entries equal to zero except for a one in the $k$th position, $k=1,\ldots, r$. Then, the binarization of $\{X_t, t \in \mathbb{Z}\}$ is given by the process $\{\bm Y_t=(Y_{t,1}, \ldots, Y_{t,r}), t \in \mathbb{Z}\}$ such that $\bm Y_t=\bm e_j$ if $X_t=j$. Fixed $l \in \mathbb{Z}$ and $i,j\in \mathcal{V}$, consider the correlation
	\begin{equation}\label{corindicators}
		\phi_{ij}(l)=Corr(Y_{t, i}, Y_{t-l, j}),
	\end{equation}
	which measures linear dependence between the $i$th and $j$th categories with respect to the lag $l$. According to Theorem 1 in \cite{lopez2023hard}, the quantity $\phi_{ij}(l)$ describes both the signed and unsigned dependence patterns of a categorical process. Moreover, this quantity can be written as
	
	\begin{equation}\label{phiijl}
		\phi_{ij}(l)=\frac{p_{ij}(l)-\pi_i\pi_j}{\sqrt{\pi_i(1-\pi_i)\pi_j(1-\pi_j)}}.
	\end{equation}

Based on previous comments, a complete description of process $X_t$ can be obtained by considering the matrix $\bm \Phi(l)=\big(\phi_{ij}(l)\big)_{1 \le i,j \le r}$. This matrix can be directly estimated by means of $\widehat{\bm \Phi}(l)=\big(\widehat{\phi}_{ij}(l)\big)_{1\le i, j\le r}$, where  the estimates $\widehat{\phi}_{ij}(l)$ are computed as
\begin{equation}\label{estimatecorrelation}
	\widehat{\phi}_{ij}(l)=\frac{\widehat{p}_{ij}(l)-\widehat{\pi}_i\widehat{\pi}_j}{\sqrt{\widehat{\pi}_i(1-\widehat{\pi}_i)\widehat{\pi}_j(1-\widehat{\pi}_j)}},
\end{equation}
with $\widehat\pi_i$ ($\widehat\pi_j$) and $\widehat p_{ij}(l)$ given in \eqref{estimates}.

Note that all the previously introduced features are well-defined for any stationary process. However, when assuming a specific type of parametric model, we can describe the process $X_t$ by means of the corresponding vector of parameters, denoted by $\boldsymbol \theta$. For instance, if $X_t$ is a Markov chain (MC), then $\boldsymbol \theta$ is given by the vectorized version of the transition probability matrix. When dealing with a realization of the process, the vector $\boldsymbol \theta$ must be estimated in a specific way, e.g., via maximum likelihood estimation (MLE), giving rise to the vector of estimated parameters $\widehat{\boldsymbol \theta}$. 

\subsection{Three distances between categorical processes}\label{subsectionthreedistances}

Hereafter, $\{X_t^{(1)}, t \in \mathbb{Z}\}$ and $\{X_t^{(2)}, t \in \mathbb{Z}\}$ (or just $X_t^{(1)}$ and $X_t^{(2)}$) denote two independent categorical stochastic processes with the same properties as the process $\{X_t, t \in \mathbb{Z}\}$ introduced above. Similarly, ${\boldsymbol X_T}^{(1)}=\big(x_1^{(1)},\ldots, x_T^{(1)}\big)$ and ${\boldsymbol{X}_T}^{(2)}=\big(x_1^{(2)},\ldots, x_T^{(2)}\big)$ denote two realizations of length $T$ from processes $X_t^{(1)}$ and $X_t^{(2)}$, respectively. In addition, the superscript $(i)$ is employed to indicate that a specific feature (estimate) is associated with process $X_t^{(i)}$ (realization ${\boldsymbol X_T}^{(i)}$), $i=1,2$. For instance, $\pi_j^{(1)}$ denotes the marginal probability for the $j$th category in process $X_t^{(1)}$, and $\widehat{\pi}_j^{(1)}$ denotes the estimate of such probability according to the realization ${\boldsymbol X_T}^{(1)}$. 

According to the model-free features introduced in Section \ref{subsectionstructuralfeatures} (see \eqref{matrixv}, \eqref{vectorkappa} and \eqref{corindicators}), and following \cite{lopez2023hard}, one can define two distance measures between categorical stochastic processes. The first metric, so-called $d_{CC}$, is based on Cramer\textquotesingle s $v$ and Cohen\textquotesingle s $\kappa$, while the second distance, denoted by $d_B$, relies on the binarization of the processes. Specifically, dissimilarities $d_{CC}$ and $d_B$ are defined as follows

\begin{equation}
	\begin{split}
	d_{CC}(X_t^{(1)}, X_t^{(2)})=\sum_{k=1}^{L}\Big[\left\lVert vec\big({\bm V}(l_k)^{(1)}-{\bm V}(l_k)^{(2)}\big)\right\rVert^2 \\
	+\, \, \left\Vert {\bm{\mathcal{K}}}(l_k)^{(1)}-{\bm{\mathcal{K}}}(l_k)^{(2)}\right\Vert^2\Big] + \left\Vert {\bm \pi}^{(1)}-{\bm \pi}^{(2)}\right\Vert^2,
	\end{split}
\end{equation}

\begin{equation}
	\begin{split}
		d_{B}(X_t^{(1)}, X_t^{(2)})= \sum_{k=1}^{L} \left\lVert vec\big({\bm \Phi}(l_k)^{(1)}-{\bm \Phi}(l_k)^{(2)}\big)\right\rVert^2  + \left\lVert {\bm \pi}^{(1)}-{\bm \pi}^{(2)} \right\rVert^2,
	\end{split}
\end{equation}

\noindent where the operator $vec(\cdot)$ transforms a matrix into a row vector by sequentially placing the corresponding numbers by columns and $\mathcal{L}=\{l_1, \ldots, l_L\}$ is a set of lags which is determined by the user. The metric $d_{CC}$ combines the features $V_{ij}(l)$ in \eqref{matrixv} with the quantities $\mathcal{K}_i(l)$ in \eqref{vectorkappa}, thus taking into account signed and unsigned dependence simultaneously. On the other hand, the distance $d_{B}$ jointly considers both types of dependence, thus evaluating discrepancy between the whole serial dependence patterns of the series. Note that a term measuring discrepancy between the marginal distributions appears in the definition of both metrics. It is worth highlighting that this term improves the discriminative ability of both dissimilarities (see Remark 4 in Section 2 of \cite{lopez2023hard}). 

Both metrics $d_{CC}$ and $d_B$ are defined under the general assumption of stationarity. An alternative way of assessing discrepancy between both processes is by assuming a common parametric model and evaluating dissimilarity between the vectors of model parameters. The corresponding metric, denoted by $d_{MLE}$, is defined as

\begin{equation}
	\begin{split}
		d_{MLE}(X_t^{(1)}, X_t^{(2)})=\left\lVert \boldsymbol \theta^{(1)} - \boldsymbol \theta^{(2)}\right\rVert^2.
	\end{split}
\end{equation}

Note that, in practice, the three dissimilarities previously introduced must be properly estimated from realizations ${\boldsymbol X_T}^{(1)}$ and ${\boldsymbol X_T}^{(2)}$, which leads to estimates of $d_{CC}$, $d_B$ and $d_{MLE}$ given by

\begin{equation}
	\begin{split}
		\widehat{d}_{CC}(X_t^{(1)}, X_t^{(2)})=\sum_{k=1}^{L}\Big[\left\lVert vec\big(\widehat{\bm V}(l_k)^{(1)}-\widehat{\bm V}(l_k)^{(2)}\big)\right\rVert^2 \\
		+\, \, \left\Vert \widehat{\bm{\mathcal{K}}}(l_k)^{(1)}-\widehat{\bm{\mathcal{K}}}(l_k)^{(2)}\right\Vert^2\Big] + \left\Vert \widehat{\bm \pi}^{(1)}-\widehat{\bm \pi}^{(2)}\right\Vert^2,
	\end{split}
\end{equation}

\begin{equation}
	\begin{split}
		\widehat{d}_{B}(X_t^{(1)}, X_t^{(2)})= \sum_{k=1}^{L} \left\lVert vec\big(\widehat{\bm \Phi}(l_k)^{(1)}-\widehat{\bm \Phi}(l_k)^{(2)}\big)\right\rVert^2  + \left\lVert \widehat{\bm \pi}^{(1)}-\widehat{\bm \pi}^{(2)} \right\rVert^2,
	\end{split}
\end{equation}

\begin{equation}
	\begin{split}
		\widehat{d}_{MLE}(X_t^{(1)}, X_t^{(2)})=\left\lVert \widehat{\boldsymbol \theta}^{(1)} - \widehat{\boldsymbol \theta}^{(2)}\right\rVert^2,
	\end{split}
\end{equation}

\noindent respectively, where $\widehat{\bm \pi}^{(i)}=(\widehat{\pi}_1^{(i)}, \ldots, \widehat{\pi}_r^{(i)})$, $i=1,2$. Distances $\widehat{d}_{CC}$, $\widehat{d}_B$ and $\widehat{d}_{MLE}$ have been used in \cite{lopez2023hard} to perform clustering of CTS. Specifically, their behavior was analyzed in a broad simulation study involving several types of categorical models, and the advantages and disadvantages of each metric were discussed. In short, metrics $\widehat{d}_{CC}$ and $\widehat{d}_{B}$ showed a better clustering effectiveness than distance $\widehat{d}_{MLE}$, even though the latter metric takes advantage of assuming the true generating mechanism, which is not realistic in practice. 

According to the form of metrics  $\widehat{d}_{CC}$, $\widehat{d}_B$ and $\widehat{d}_{MLE}$ and the null hypothesis in \eqref{hypothesis1}, a reasonable decision rule would rely on rejecting this hypothesis for large values of the considered distances. To that aim, a proper approximation of the null distribution of these metrics is needed. 
	
\section{Bootstraps tests for categorical series}\label{sectionbootstraptests}
	
		Bootstrap methods provide a powerful way of approximating the null distribution of distances $\widehat{d}_{CC}$, $\widehat{d}_{B}$ and $\widehat{d}_{MLE}$. In this section, three resampling procedures based on bootstrapping these metrics are proposed. The first test is a parametric method based on generating bootstrap replicates by considering the average vector of estimated model coefficients via maximum likelihood. The remaining two approaches rely on well-known resampling methods for dependent data. The key principle is to draw pseudo-series capturing the dependence structure without assuming any parametric model. It is worth highlighting that the proposed bootstrap approaches have already been considered by \cite{lopez2021quantile} in a context of multivariate time series. 
	
	\subsection{A test based on estimated model coefficients}
	
	The first test we propose is a parametric procedure. Specifically, for the $T$-length realizations ${\boldsymbol X_T}^{(1)}=\big(x_1^{(1)},\ldots, x_T^{(1)}\big)$ and ${\boldsymbol{X}_T}^{(2)}=\big(x_1^{(2)},\ldots, x_T^{(2)}\big)$, and a distance measure between CTS, $\widehat{d}\in\{\widehat{d}_{CC}, \widehat{d}_{B}, \widehat{d}_{MLE}\}$, the method is based on the following steps:
	
	\vspace*{0.25cm}
	
	\noindent \textsc{Step 1}. Select a specific class of categorical model (e.g., a MC of order 1).
	
	\vspace*{0.15cm}
	
	\noindent \textsc{Step 2}. For each one of the realizations ${\boldsymbol X_T}^{(1)}$ and ${\boldsymbol X_T}^{(2)}$, estimate via maximum likelihood the vector of parameters for the categorical model selected in the previous step, which results in the vectors $\widehat{\bm \theta}^{(1)}$ and $\widehat{\bm \theta}^{(2)}$, respectively. Compute the vector of average estimates as $\widetilde{\bm \theta}=\frac{\widehat{\bm \theta}^{(1)}+\widehat{\bm \theta}^{(2)}}{2}$.   
	
	\vspace*{0.15cm}
	
	\noindent \textsc{Step 3}. Simulate two independent time series of length $T$, ${\bm X_T}^{(1)*}$ and ${\bm X_T}^{(2)*}$, by considering the categorical model selected in the first step with parameters given by $\widetilde{\bm \theta}$. Then, obtain the bootstrap version $\widehat{d}^*$ of $\widehat{d}$ based on the series ${\boldsymbol X_T}^{(1)*}$ and ${\boldsymbol X_T}^{(2)*}$.
	
	\vspace*{0.15cm}

	\noindent \textsc{Step 4}. Repeat Step~3 a large number $B$ of times to obtain the bootstrap replicates $\widehat{d}^{(1)*}, \ldots, \widehat{d}^{(B)*}$.
	
	\vspace*{0.15cm}
	
	\noindent \textsc{Step 5}. Given a significance level $\alpha$, compute the quantile of order $1-\alpha$ based on the sample $\widehat{d}^{\,(1)*}, \ldots, \widehat{d}^{\,(B)*}$, denoted by $q_{1-\alpha}^*$. Then, the decision rule consists of rejecting $H_0$ if $\widehat{d}(X_t^{(1)}, X_t^{(2)})>q_{1-\alpha}^*$. 
	
	Note that the consideration of the average vector $\widetilde{\bm \theta}$ in the previous procedure allows for a proper approximation of the distribution of $\widehat{d}$ under the null hypothesis independently of this hypothesis being true.
	
	From now on, we will refer to the test presented in this section as bootstrap averaging (BA).  
	
	\subsection{A test based on the moving blocks bootstrap}
	
	In this section, we introduce an alternative bootstrap test based on a modi\-fi\-ca\-tion of the classical MBB method proposed by \cite{kunsch1989jackknife} and \cite{liu1992moving}. MBB generates replicates of the time series by joining blocks of fixed length which have been drawn randomly with replacement from among blocks of the original realizations. This approach allows to mimic the underlying dependence structure without assuming specific parame\-tric models for the generating processes.
	
	Given the realizations ${\boldsymbol X_T}^{(1)}=\big(x_1^{(1)},\ldots, x_T^{(1)}\big)$ and ${\boldsymbol{X}_T}^{(2)}=\big(x_1^{(2)},\ldots, x_T^{(2)}\big)$, and a distance measure between CTS, $\widehat{d}\in\{\widehat{d}_{CC}, \widehat{d}_{B}, \widehat{d}_{MLE}\}$, the procedure proceeds as follows:
	
	\vspace*{0.25cm}
	
	\noindent \textsc{Step 1}. Fix a positive integer, $b$, representing the block size, and take $k$ equal to the smallest integer greater than or equal to $T/b$. 
	\vspace*{0.15cm}
	
	\noindent \textsc{Step 2}. For each realization ${\boldsymbol X_T}^{(i)}$, define the block $\bm B_j^{(i)}=\big(x_j^{(i)}, \ldots, x_{j+b-1}^{(i)}\big)$, for $j=1,\ldots,q$, with $q=T-b+1$. Let $\overline{\bm B}=\{\bm B_j^{(1)}, \ldots, \bm B_q^{(1)}, \bm B_j^{(2)}, \ldots, \bm B_q^{(2)}\}$ be the set of all blocks, those coming from  ${\bm X_T}^{(1)}$ and those coming from ${\bm X_T}^{(2)}$. 
	
	\vspace*{0.15cm}
	
	\noindent \textsc{Step 3}. Draw two sets of $k$ blocks, $\boldsymbol \xi^{(i)}= \big(\boldsymbol \xi^{(i)}_1, \ldots, \boldsymbol \xi^{(i)}_k\big)$, $i=1,2$, with equiprobable distribution from $\overline {\boldsymbol B}$. Note that each $\boldsymbol \xi^{(i)}_j$, $j=1,\ldots,k$, $i=1,2$, is a $b$-length CTS, let us say $( \xi^{(i)}_{1j}, \xi^{(i)}_{2j}, \ldots, \xi^{(i)}_{bj})$. 
	
	\vspace*{0.15cm}
	
	\noindent \textsc{Step 4}. For each $i=1,2$, construct the pseudo-series ${\boldsymbol X_T}^{(i)*}$ by taking the first $T$ elements of:
	$$
		\boldsymbol \xi^{(i)}= (\xi^{(i)}_{11}, \xi^{(i)}_{21}, \ldots, \xi^{(i)}_{b1}, \xi^{(i)}_{12}, \xi^{(i)}_{22}, \ldots, \xi^{(i)}_{b2}, \ldots, \xi^{(i)}_{1k}, \xi^{(i)}_{2k}, \ldots, \xi^{(i)}_{bk}).
	$$
	Then, obtain the bootstrap version $\widehat{d}^*$ of $\widehat{d}$ based on the pseudo-series ${\boldsymbol X_T}^{(1)*}$ and ${\boldsymbol X_T}^{(2)*}$. 
	
	\vspace*{0.15cm}
	
	\noindent \textsc{Step 5}. Repeat Steps~3 and 4 a large number $B$ of times to obtain the bootstrap replicates $\widehat{d}^{(1)*}, \ldots, \widehat{d}^{(B)*}$.
	
	\vspace*{0.15cm}
	
	\noindent \textsc{Step 6}. Given a significance level $\alpha$, compute the quantile of order $1-\alpha$ based on the sample $\widehat{d}^{\,(1)*}, \ldots, \widehat{d}^{\,(B)*}$, denoted by $q_{1-\alpha}^*$. Then, the decision rule consists of rejecting $H_0$ if $\widehat{d}(X_t^{(1)}, X_t^{(2)})>q_{1-\alpha}^*$. 
	
	\vspace*{0.15cm}
	
	Note that, by considering the whole set of blocks $\overline {\bm B}$ in Step 2, both pseudo-time series ${\bm X_T}^{(1)*}$ and ${\bm X_T}^{(2)*}$ are expected to contain information about the original series ${\bm X_T}^{(1)}$ and ${\bm X_T}^{(2)}$ in equal measure. This way, the bootstrap procedure is able to correctly approximate the distribution of the test statistic $\widehat{d}$ under the null hypothesis even if this hypothesis is not true. 
	
	From now on, we will refer to the test presented in this section as MBB. 
	
	\subsection{A test based on the stationary bootstrap}
	
	The third bootstrap mechanism to approximate the distribution of $\widehat{d}\in\{\widehat{d}_{CC}, \widehat{d}_{B}, \widehat{d}_{MLE}\}$ adapts the classical SB \cite{politis1994stationary}. This resampling method is aimed at overcoming the lack of stationarity of the MBB procedure. Note that $d_{CC}$ and $d_B$ are well-defined only for stationary processes, so it is desirable that a bootstrap technique based on estimates of these metrics generates stationary pseudo-series.  
	
	For realizations ${\boldsymbol X_T}^{(1)}=\big(x_1^{(1)},\ldots, x_T^{(1)}\big)$ and ${\boldsymbol{X}_T}^{(2)}=\big(x_1^{(2)},\ldots, x_T^{(2)}\big)$, the resampling method proceeds as follows:
	
	\vspace*{0.25cm}
	
	\noindent \textsc{Step 1}. Fix a real number $p \in [0,1]$.
	
	\vspace*{0.15cm}
	
	\noindent \textsc{Step 2}. For $i=1,2$, draw randomly one observation from the pooled series $\widetilde {\boldsymbol X}=\big({\boldsymbol X_T}^{(1)}, {\boldsymbol X_T}^{(2)}\big)$. The drawn observations are of the form $x_{j^i}^{(k^i)}$ for some $k^i=1,2$, $j^i=1,\ldots,T$, and $i=1,2$. Then, $x_{j^i}^{(k^i)}$ is taken as the first element of the pseudo-series ${\boldsymbol X_T}^{(i)*}$, denoted by ${x_1}^{(i)*}$.
	
	\vspace*{0.15cm}
	
	\noindent \textsc{Step 3}. Once obtained ${x_l}^{(i)*}=x_{j^i}^{(k^i)}$, for $l<T$ and $i=1,2$, the next bootstrap replication ${x}_{l+1}^{(i)*}$ is defined as $x_{j^i+1}^{(k^i)}$ with probability $1-p$, and is randomly drawn from $\widetilde {\boldsymbol X}$ with probability $p$. When $j^i=T$, the selected observation is $x_{1}^{(2)}$ if $k^i=1$ and $x_{1}^{(1)}$ if $k^i=2$.
	
	\vspace*{0.15cm}
	
	\noindent \textsc{Step 4}. Repeat Step 3 until the pseudo-series ${\boldsymbol X_T}^{(1)*}$ and ${\boldsymbol X_T}^{(2)*}$ contain $T$ observations. Based on these pseudo-series, compute the bootstrap version $\widehat{d}^*$ of $\widehat{d}$. 
	
	\vspace*{0.15cm}
	
	\noindent \textsc{Step 5}. Repeat Steps~2-4 a large number $B$ of times to obtain the bootstrap replicates $\widehat{d}^{(1)*}, \ldots, \widehat{d}^{(B)*}$.
	
	\vspace*{0.15cm}
	
	\noindent \textsc{Step 6}. Given a significance level $\alpha$, compute the quantile of order $1-\alpha$ based on the sample $\widehat{d}^{\,(1)*}, \ldots, \widehat{d}^{\,(B)*}$, denoted by $q_{1-\alpha}^*$. Then, the decision rule consists of rejecting $H_0$ if $\widehat{d}(X_t^{(1)}, X_t^{(2)})>q_{1-\alpha}^*$. 
	
	\vspace*{0.15cm}
	
	It is worth remarking that, likewise MBB procedure, a proper approximation of the null distribution of $\widehat{d}$ is also expected here due to the consideration of the pooled time series $\widetilde {\boldsymbol X}$ in the generating mechanism.
	
	From now on, we will refer to the test presented in this section as SB. 
	
	\section{Simulation study}\label{sectionsimulationstudy}
	
	In this section, we carry out a simulation study conducted to assess the performance with finite samples of the testing procedures presented in Section \ref{sectionbootstraptests}. Note that we are considering three dissimilarities and three resampling schemes, which gives rise to 9 hypothesis tests to be evaluated. First we describe the simulation mechanism and then we discuss the main results. Finally, some additional analysis are performed to analyze the procedures in deeper detail. 
	
	\subsection{Experimental design}\label{subsectioned}
	
	The behavior of the methods was examined with pairs of CTS realizations, ${\boldsymbol X_T}^{(1)}=\big(x_1^{(1)},\ldots, x_T^{(1)}\big)$ and ${\boldsymbol X_T}^{(2)}=\big(x_1^{(2)},\ldots, x_T^{(2)}\big)$, simulated from categorical processes selected to cover different dependence structures. Specifically, three types of generating models were considered, namely MC, HMM, and new discrete ARMA (NDARMA) processes. In all cases, the deviation from the null hypothesis in \eqref{hypothesis1} was established in accordance with differences in the coefficients of the generating processes. Specifically, the degree of deviation between the simulated realizations was regulated by a specific parameter ($\delta$) included in the formulation of the models. The specific scenarios and generating processes are given below.  
	
	\vspace*{0.15cm}
	
	\noindent \textbf{Scenario 1}. Hypothesis testing for MC. Consider three-state MC models of order 1 given by the matrix of transition probabilities
	\begin{equation*}
		\begin{pmatrix}
			0.1+\delta & 0.1+\delta & 0.1+\delta  \\
			0.3+\delta & 0.3+\delta  & 0.3+\delta  \\
			0.6-2\delta & 0.6-2\delta & 0.6-2\delta
		\end{pmatrix}.
	\end{equation*}
	
	\vspace*{0.15cm}
	
	\noindent \textbf{Scenario 2}. Hypothesis testing for HMM. Consider three-state HMM models of order 1 defined by the same transition and emission probability matrix, which is given by
	\begin{equation*}
		\begin{pmatrix}
			0.3+\delta & 0.3+\delta & 0.3+\delta  \\
			0.3+\delta & 0.3+\delta  & 0.3+\delta  \\
			0.4-2\delta & 0.4-2\delta & 0.4-2\delta
		\end{pmatrix}.
	\end{equation*}
	
	\vspace*{0.15cm}
	
	\noindent \textbf{Scenario 3}. Hypothesis testing for NDARMA models. Let $\{X_t, t \in \mathbb{Z}\}$ and $\{\epsilon_t, t \in \mathbb{Z}\}$ be two count processes with range $\{1,\ldots,r\}$ and following the equation
	\begin{equation}
		\begin{split}
			X_t=\alpha_{t,1}X_{t-1}+\ldots+\alpha_{t,p}X_{t-p}+
			\beta_{t,0}\epsilon_{t}+\ldots+\beta_{t,q}\epsilon_{t-q},
		\end{split}
	\end{equation}
	where $\{\epsilon_t, t \in \mathbb{Z}\}$ is i.i.d. with $P(\epsilon_t=i)=\pi_i$, independent of $(X_s)_{s <t}$, and the i.i.d. multinomial random vectors 
	\begin{equation}
		\begin{split}
			(\alpha_{t,1}, \ldots, \alpha_{t,p}, \beta_{t,0},\ldots, \beta_{t,q}) \sim  \text{MULT}(1;\phi_1, \ldots, \phi_p,\varphi_0, \ldots, \varphi_q),
		\end{split}
	\end{equation}
	are independent of $\{\epsilon_t, t \in \mathbb{Z}\}$ and $(X_s)_{s <t}$. The considered models are three-state NDARMA(1,0) processes with marginal probabilities given by the vector $(\pi_1, \pi_2, \pi_3)=(0.2, 0.3-\delta, 0.5+\delta)$ and multinomial probabilities given by the vector $(\phi_1, \varphi_0)=(0.6-\delta, 0.4+\delta)$.     
	
	\vspace*{0.20cm}
	
In the previous scenarios, ${\boldsymbol X_T}^{(1)}$ is always generated by taking $\delta=0$, while ${\boldsymbol X_T}^{(2)}$ is generated using different values of $\delta$, thus allowing to obtain simulation schemes under the null, when $\delta=0$ also for ${\boldsymbol X_T}^{(2)}$, and under the alternative otherwise. To empirically assess the size and power behavior of the different tests, a number of $N=1000$ replications of pairs of realizations ${\boldsymbol X_T}^{(1)}$ and 
${\boldsymbol X_T}^{(2)}$ coming from the processes at each scenario were obtained. Realizations ${\boldsymbol X_T}^{(2)}$ were generated by considering $\delta \in \{0, 0.05, 0.075, 0.10\}$, $\{0, 0.025, 0.05, 0.075\}$ and $\{0, 0.10, 0.15, 0.20\}$ in Scenarios 1, 2 and 3, respectively. 

For a pair of realizations associated with a specific value of $\delta$, $B=500$ bootstrap replicates were considered to approximate the distribution of the different test statistics under the null hypothesis. Simulations were carried out for different values of $T$, namely $T\in\{100, 200, 500\}$. For methods MBB and SB, we chose the corresponding input parameters as $b=\ceil{T^{1/3}}$ and $p=T^{-1/3}$, respectively, being $\ceil{\cdot}$ the ceiling function. These choices were motivated by the related literature. For instance, \cite{hall1995blocking} addressed the issue of selecting $b$ in the context of bias and variance bootstrap estimation, concluding that the optimal block size is of order $T^{1/3}$. On the other hand, since the mean block size in SB corresponds to $1/p$, it is reasonable to select $p$ of order $T^{-1/3}$. Computation of dissimilarities $\widehat{d}_{CC}$ and $\widehat{d}_B$ was carried out by considering only one lag, i.e, $\mathcal{L}=\{1\}$, since the first lag is enough to characterize the dependence structures of the processes in the three scenarios. Computation of distance $\widehat{d}_{MLE}$ was performed by considering the true class of models existing in each scenario. Note that, for each combination of metric and resampling scheme, each one of the $N$ replications leads to a particular outcome of the decision rule for the test in \eqref{hypothesis1}. In all cases, the results were obtained for a significance level $\alpha=0.05$.

	\subsection{Results and discussion}\label{subsectionrd}
	
	Tables \ref{tablescenario1}, \ref{tablescenario2} and \ref{tablescenario3} contain the rejection rates for Scenarios 1, 2 and 3, respectively. In Scenario 1 and, under the null hypothesis ($\delta=0$), all methods display rejection rates slightly above the significance level (0.05) when $T=100$. However, when increasing the series length, the rates get close to this level. In fact, for $T=500$, all approaches approximate the nominal size pretty well, with the tests based on the distance $\widehat{d}_{MLE}$ being slightly conservative. On the other hand, when the null hypothesis is not true ($\delta>0$), there are dramatic differences in the rejection rates of the considered approaches. Specifically, metric $\widehat{d}_{CC}$ achieves the highest power by a large degree, while distance $\widehat{d}_{B}$ gets very poor results. Metric $\widehat{d}_{MLE}$ lies somewhere in the middle. For a given dissimilarity, there are no substantial differences among the three bootstrap techniques, although the MBB approach slightly produces the highest rejection rates in most of the settings with $\delta>0$. As expected, all methods improve their power when increasing the value of the parameter $\delta$ and the series length. 
	
	In Scenario 2 (see Table \ref{tablescenario2}), the metric $\widehat{d}_{CC}$ exhibits again the largest power, but the differences between the considered approaches are less substantial. As in Scenario 1, the method based on moving blocks (MBB) moderately outperforms the remaining resampling techniques. Finally, in Scenario 3 (see Table \ref{tablescenario3}), the results are quite similar to the ones in Scenario 1, with the metric $\widehat{d}_{CC}$ clearly outperforming the remaining dissimilarities in most cases.
	
	In short, the above simulation results showed that, under the null hypothesis, most methods respect the significance level rather properly when sufficiently large values of the series length are considered. On the other hand, when the null hypothesis is not true, the test based on the metric $\widehat{d}_{CC}$ and the bootstrap approach MBB exhibits the highest power in most settings. Note that this fact is quite interesting and advantageous for the practitioners, since neither the dissimilarity $\widehat{d}_{CC}$ nor the resampling mechanism based on moving blocks assume a specific class of categorical models. Moreover, as stated in Section \ref{subsectioned}, the rejection rates provided in Tables \ref{tablescenario1}, \ref{tablescenario2} and \ref{tablescenario3} were obtained by considering the default value $b=\ceil{T^{1/3}}$ for the block size in all cases, which means that no hyperparameter selection was performed for MBB.

	\begin{table}
		\centering
		\resizebox{12cm}{!}{\begin{tabular}{c|ccc|ccc|ccc}\hline 
			  &  & $T=100$ &  &  & $T=200$ &  &  & $T=500$ &  \\
			  &  BA & MBB & SB & BA & MBB & SB & BA & MBB & SB  \\ \hline
		  $\delta=0.000$	  & & & & & & & & & \\  
		  $\widehat{d}_{CC}$   & 0.058 & 0.067 & 0.069 & 0.056 & 0.066 & 0.064 & 0.053 & 0.054 & 0.052 \\ 
		  $\widehat{d}_{B}$  & 0.067 & 0.084 & 0.086 & 0.057 & 0.072 & 0.060 & 0.054 & 0.060 & 0.053 \\ 
		  $\widehat{d}_{MLE}$  & 0.054 & 0.059 & 0.067 & 0.050 & 0.053 & 0.063 & 0.036 & 0.044 & 0.043 \\   \hline 
		  $\delta=0.050$	  & & & & & & & & & \\  
		  $\widehat{d}_{CC}$   & 0.242 & 0.313 & 0.307 & 0.411 & 0.454 & 0.410 & 0.823 & 0.854 & 0.820 \\
		  $\widehat{d}_{B}$   & 0.062 & 0.078 & 0.079 & 0.059 & 0.082 & 0.083 & 0.102 & 0.121 & 0.114 \\ 
		  $\widehat{d}_{MLE}$  & 0.125 & 0.135 & 0.101 & 0.179 & 0.202 & 0.154 & 0.461 & 0.496 & 0.353 \\ \hline 
		  $\delta=0.075$	  & & & & & & & & & \\  
		  $\widehat{d}_{CC}$   &  0.474& 0.523 & 0.501 & 0.745 & 0.799 & 0.752 & 0.995 & 0.993 & 0.996 \\
		  $\widehat{d}_{B}$ & 0.089  & 0.112 & 0.121 & 0.095  & 0.127 & 0.123 & 0.275  & 0.282 & 0.265 \\
		  $\widehat{d}_{MLE}$  &  0.227 & 0.249 & 0.166 & 0.406 & 0.436 & 0.283 & 0.886 & 0.899 & 0.749 \\ \hline 
		  $\delta=0.100$	  & & & & & & & & & \\  
		  $\widehat{d}_{CC}$  & 0.720 & 0.763 & 0.715 & 0.963 & 0.972 & 0.961 & 1.000 & 1.000 & 1.000 \\
		  $\widehat{d}_{B}$ & 0.196 & 0.208 & 0.222 & 0.508 & 0.546 & 0.513 & 0.996 & 0.997 & 0.996 \\ 
		  $\widehat{d}_{MLE}$ & 0.425 & 0.459 & 0.263 & 0.744 & 0.783 & 0.555 & 1.000 & 1.000 & 0.991 \\ \hline 
		\end{tabular}}
	\caption{Simulated rejection rates in Scenario 1.}
	\label{tablescenario1}
	\end{table}

\begin{table}
	\centering
	\resizebox{12cm}{!}{\begin{tabular}{c|ccc|ccc|ccc}\hline 
		&  & $T=100$ &  &  & $T=200$ &  &  & $T=500$ &  \\
		&  BA & MBB & SB & BA & MBB & SB & BA & MBB & SB  \\ \hline
			$\delta=0.000$	  & & & & & & & & & \\  
		$\widehat{d}_{CC}$   & 0.061 & 0.057 & 0.045 & 0.050 & 0.056 & 0.068 & 0.051 & 0.057 & 0.053 \\ 
		$\widehat{d}_{B}$  & 0.055 & 0.084 & 0.071 & 0.055 & 0.072 & 0.062 & 0.054 & 0.073 & 0.055 \\ 
		$\widehat{d}_{MLE}$  & 0.063 & 0.055 & 0.071 & 0.050 & 0.055 & 0.052 & 0.056 & 0.054 & 0.047 \\   \hline 
		$\delta=0.025$	  & & & & & & & & & \\  
		$\widehat{d}_{CC}$   & 0.161  & 0.203  & 0.188 & 0.255  & 0.315  & 0.262 & 0.373  &  0.452  & 0.340  \\
		$\widehat{d}_{B}$ & 0.071  & 0.089  & 0.081 & 0.145  & 0.176  & 0.147 & 0.201  &  0.252  & 0.234  \\
		$\widehat{d}_{MLE}$ & 0.103  & 0.182  & 0.150 & 0.216  & 0.293  & 0.223 & 0.313  &  0.312  & 0.326  \\ \hline 
		$\delta=0.050$	  & & & & & & & & & \\  
		$\widehat{d}_{CC}$   & 0.267  & 0.294  & 0.280 & 0.357  & 0.389  & 0.342& 0.593  &  0.701  & 0.650  \\
		$\widehat{d}_{B}$ & 0.145  & 0.176  & 0.134 & 0.287  & 0.298  & 0.284& 0.465  &  0.523  & 0.434  \\
		$\widehat{d}_{MLE}$ & 0.228  & 0.244  & 0.256 & 0.327  & 0.346  & 0.312& 0.591  &  0.673  & 0.595  \\ \hline 
		$\delta=0.075$	  & & & & & & & & & \\  
		$\widehat{d}_{CC}$   & 0.404  & 0.452  & 0.431 & 0.661  & 0.705  & 0.654 & 0.843  &  0.964  & 0.875  \\
		$\widehat{d}_{B}$ & 0.268  & 0.297  & 0.259 & 0.476  & 0.513  & 0.487 & 0.712  &  0.779  & 0.734 \\   
		$\widehat{d}_{MLE}$& 0.358  & 0.417  & 0.401 & 0.585  & 0.685  & 0.624 & 0.813  &  0.924  & 0.825  \\  \hline 
	\end{tabular}}
	\caption{Simulated rejection rates in Scenario 2.}
	\label{tablescenario2}
\end{table}

\begin{table}
	\centering
	\resizebox{12cm}{!}{\begin{tabular}{c|ccc|ccc|ccc}\hline 
		&  & $T=100$ &  &  & $T=200$ &  &  & $T=500$ &  \\
		&  BA & MBB & SB & BA & MBB & SB & BA & MBB & SB  \\ \hline
		$\delta=0.00$	  & & & & & & & & & \\  
		$\widehat{d}_{CC}$   & 0.081 & 0.077 & 0.067 & 0.042 & 0.039 & 0.042 & 0.045 & 0.046 & 0.042 \\ 
		$\widehat{d}_{B}$  & 0.071 & 0.091 & 0.081 & 0.054 & 0.058 & 0.062 & 0.048 & 0.048 & 0.052 \\ 
		$\widehat{d}_{MLE}$ & 0.060 & 0.074 & 0.070 & 0.045 & 0.062 & 0.051 & 0.057 & 0.060 & 0.051 \\ \hline 
		$\delta=0.10$	  & & & & & & & & & \\  
		$\widehat{d}_{CC}$   &  0.341 & 0.407 & 0.335 & 0.640 & 0.667  & 0.623 & 0.998  & 0.997  & 0.997  \\ 
		$\widehat{d}_{B}$ & 0.053 & 0.076 & 0.072 & 0.067 & 0.093 & 0.089& 0.143 & 0.163  & 0.148 \\
		$\widehat{d}_{MLE}$ & 0.179  & 0.206 & 0.196 & 0.331 & 0.366 & 0.345 & 0.760 & 0.793 & 0.771 \\ \hline 
		$\delta=0.15$	  & & & & & & & & & \\  
		$\widehat{d}_{CC}$   & 0.665 & 0.700 & 0.625 & 0.915 & 0.920 & 0.905 & 1.000 & 1.000 & 1.000 \\ 
		$\widehat{d}_{B}$ & 0.078 & 0.093 & 0.080 & 0.126 & 0.140 & 0.123 & 0.390 & 0.410 & 0.405 \\
		$\widehat{d}_{MLE}$ & 0.370 & 0.424 & 0.404 & 0.673 & 0.701 & 0.678 & 0.991 & 0.996 & 0.996 \\ \hline 
		$\delta=0.20$	  & & & & & & & & & \\  
	    $\widehat{d}_{CC}$    & 0.915 & 0.925 & 0.925 & 1.000 & 1.000 & 1.000 & 1.000 & 1.000 & 1.000 \\ 
	    $\widehat{d}_{B}$ & 0.124 & 0.126 & 0.143 & 0.260 & 0.292 & 0.267 & 0.800 & 0.822 & 0.820 \\ 
	    $\widehat{d}_{MLE}$ & 0.619 & 0.659 & 0.665 & 0.940 & 0.945 & 0.945 & 1.000 & 1.000 & 1.000 \\ \hline 
	\end{tabular}}
	\caption{Simulated rejection rates in Scenario 3.}
	\label{tablescenario3}
\end{table}
	
	\subsection{Further analysis}\label{subsectionfa}
	
	In order to provide a more comprehensive evaluation of the proposed clustering methods, we extended the previous simulations by: (i) increasing the complexity of original Scenarios 1, 2 and 3, and (ii) analyzing the impact that parameters $b$ and $p$ have on the behavior of MBB and SB, respectively. Each one of the above points is discussed below. 
	
	\subsubsection{Additional scenarios}
	
	Two additional setups were constructed by increasing the complexity of Scenarios 1, 2 and 3. First, note that the series range in these scenarios was fixed to $\mathcal{V}=\{1, 2, 3\}$. However, it is interesting to assess the performance of the different methods when the set $\mathcal{V}$ contains a different number of categories. To this aim, we constructed a new simulation scenario, so-called Scenario 4, in which the size of $\mathcal{V}$ is randomly determined. Specifically, let $R$ be a random variable following a discrete uniform distribution in the set $\{2, 3, 4, 5\}$ and consider $R$-state MC models given by the following transition probability matrix of order $R$:
	
	\begin{equation}
	\begin{pmatrix} 
		\frac{1}{R}-\delta & \frac{1}{R}-\delta   &  \dots  & \frac{1}{R}-\delta  & \frac{1}{R}-\delta  \\
		\frac{1}{R} & \frac{1}{R}   &  \dots  & \frac{1}{R}  & \frac{1}{R}  \\
		\vdots & \vdots & \ddots & \vdots & \vdots \\
		\frac{1}{R} & \frac{1}{R}   &  \dots  & \frac{1}{R}  & \frac{1}{R}  \\
		\frac{1}{R}+\delta & \frac{1}{R}+\delta   &  \dots  & \frac{1}{R}+\delta  & \frac{1}{R}+\delta  \\
	\end{pmatrix}. 
	\end{equation}

The simulations concerning Scenario 4 were carried out in an analogous way as the ones described in Section \ref{subsectioned} but considering $\delta \in \{0, 0.05, 0.10, 0.15\}$. The corresponding rejection rates are displayed in Table \ref{tablescenario4}. Under the null hypothesis ($\delta=0$), all methods respect the significance level quite properly for $T=500$, while the methods based on $\widehat{d}_{CC}$ and $\widehat{d}_{B}$ show a few overrejections for $T\in\{100, 200\}$. On the other hand, the results for $\delta>0$ are rather different to the ones in Tables \ref{tablescenario1}, \ref{tablescenario2} and \ref{tablescenario3}. Dissimilarity $\widehat{d}_{B}$ still reaches the worst results by far but, this time, there seems to be no significant differences between distances $\widehat{d}_{CC}$ and $\widehat{d}_{MLE}$ in most settings. In fact, a more detailed analysis indicates that these distances get similar rejection rates for all values of $R$.  
	
	\begin{table}
		\centering
		\resizebox{12cm}{!}{\begin{tabular}{c|ccc|ccc|ccc}\hline 
				&  & $T=100$ &  &  & $T=200$ &  &  & $T=500$ &  \\
				&  BA & MBB & SB & BA & MBB & SB & BA & MBB & SB  \\ \hline
				$\delta=0.00$	  & & & & & & & & & \\  
				$\widehat{d}_{CC}$   &  0.091  & 0.067  & 0.057 & 0.074  & 0.063  & 0.054 & 0.053 & 0.057  & 0.052  \\
				$\widehat{d}_{B}$ & 0.062 & 0.064 & 0.067 & 0.060 & 0.066 & 0.070 & 0.055 & 0.053 & 0.062 \\
				$\widehat{d}_{MLE}$ & 0.048  & 0.047 & 0.058 & 0.048 & 0.049  & 0.053 & 0.051 & 0.048 & 0.052  \\ \hline 
				$\delta=0.05$	  & & & & & & & & & \\  
				$\widehat{d}_{CC}$   &  0.101 & 0.095   & 0.112  & 0.203   & 0.214   & 0.176  & 0.334 & 0.375   & 0.331   \\
				$\widehat{d}_{B}$ &  0.065 & 0.056 & 0.065 & 0.056 &  0.098 & 0.060 & 0.074 & 0.092  & 0.073  \\
				$\widehat{d}_{MLE}$ &  0.134 & 0.137  & 0.132 &  0.178 & 0.175  & 0.194 & 0.346 & 0.320  & 0.317  \\ \hline 
				$\delta=0.10$	  & & & & & & & & & \\  
				$\widehat{d}_{CC}$   & 0.194  & 0.276  & 0.234  & 0.443  & 0.470  & 0.487  & 0.843 &  0.827 & 0.804  \\
				$\widehat{d}_{B}$ & 0.104  & 0.093  & 0.091  & 0.125  & 0.122  & 0.112  & 0.151  & 0.149  & 0.152  \\
				$\widehat{d}_{MLE}$ & 0.187  & 0.273  & 0.225  & 0.364  & 0.465  & 0.451  & 0.801 & 0.824  & 0.793  \\ \hline 
				$\delta=0.15$	  & & & & & & & & & \\  
				$\widehat{d}_{CC}$   & 0.443  & 0.437  & 0.478 &  0.836 & 0.889  & 0.892 & 1.000 &  1.000 & 1.000  \\
				$\widehat{d}_{B}$ &  0.153 & 0.163 & 0.139 &  0.165 & 0.223 & 0.193 & 0.324  & 0.342  & 0.320  \\
				$\widehat{d}_{MLE}$ &  0.463 & 0.454 & 0.471  & 0.825 &  0.876 & 0.864  & 0.998 & 0.997  & 0.999  \\ \hline 
		\end{tabular}}
		\caption{Simulated rejection rates in Scenario 4.}
		\label{tablescenario4}
	\end{table}

A second additional scenario was constructed to examine the behavior of the methods when higher order dependencies exist. In particular, the so-called Scenario 5 considers three-state NDARMA(2, 0) models with marginal probabilities $(\pi_1, \pi_2, \pi_3)=(0.3, 0.3-\delta, 0.4+\delta)$ and multinomial probabilities $(\phi_1, \phi_2, \varphi_0)=(0.4-\delta, 0.4-\delta, 0.2 + 2\delta)$. Simulations were carried out this time in the same way as in previous analyses but setting $\delta \in \{0.025, 0.050, 0.075\}$ and $\mathcal{L}=\{1, 2\}$ for the computation of dissimilarities $\widehat{d}_{CC}$ and $\widehat{d}_B$, since the serial dependence structures of the processes in Scenario 5 are characterized by means of the first two lags. The corresponding rejection rates are provided in Table \ref{tablescenario5}. Once again, all methods respect the nominal size rather properly when $T=500$. Dissimilarity $\widehat{d}_{CC}$ exhibits by far the best power, and the bootstrap method MBB slightly outperforms the remaining ones in most cases. 

\begin{table}
	\centering
	\resizebox{12cm}{!}{\begin{tabular}{c|ccc|ccc|ccc}\hline 
			&  & $T=100$ &  &  & $T=200$ &  &  & $T=500$ &  \\
			&  BA & MBB & SB & BA & MBB & SB & BA & MBB & SB  \\ \hline
			$\delta=0.000$	  & & & & & & & & & \\  
			$\widehat{d}_{CC}$   &  0.047 & 0.068   & 0.062 & 0.049  & 0.068  & 0.065 & 0.061 & 0.051  & 0.054  \\
			$\widehat{d}_{B}$ & 0.070 & 0.085  & 0.084 & 0.068 & 0.064  & 0.067 & 0.049 & 0.053  & 0.054 \\
			$\widehat{d}_{MLE}$ & 0.048 & 0.050 & 0.072 & 0.049 & 0.060 & 0.058 & 0.038 & 0.048 & 0.043 \\ \hline 
			$\delta=0.025$	  & & & & & & & & & \\  
			$\widehat{d}_{CC}$   &  0.356 & 0.397 & 0.365 & 0.644 & 0.698  & 0.653 & 0.923  & 0.947  & 0.910  \\ 
			$\widehat{d}_{B}$ & 0.058 & 0.071 & 0.068 & 0.087 & 0.103 & 0.091& 0.175 & 0.193  & 0.131 \\
			$\widehat{d}_{MLE}$ &  0.174 & 0.147  & 0.126 &  0.254 & 0.307  & 0.259 & 0.373 & 0.389  & 0.293  \\ \hline 
			$\delta=0.050$	  & & & & & & & & & \\  
		$\widehat{d}_{CC}$   & 0.687 & 0.723 & 0.617 & 0.845 & 0.893 & 0.865 & 0.943 & 0.965 & 0.957 \\ 
		$\widehat{d}_{B}$ & 0.107 & 0.113 & 0.089 & 0.146 & 0.155 & 0.139 & 0.460 & 0.478 & 0.437 \\
		$\widehat{d}_{MLE}$ & 0.378 & 0.494 & 0.454 & 0.705 & 0.747 & 0.699 & 0.901 & 0.909 & 0.885 \\ \hline 
		$\delta=0.075$	  & & & & & & & & & \\  
		$\widehat{d}_{CC}$    & 0.896 & 0.915 & 0.901 & 1.000 & 1.000 & 1.000 & 1.000 & 1.000 & 1.000 \\ 
		$\widehat{d}_{B}$ & 0.145 & 0.159 & 0.153 & 0.227 & 0.302 & 0.259 & 0.805 & 0.846 & 0.819 \\ 
		$\widehat{d}_{MLE}$ & 0.627 & 0.661 & 0.654 & 0.876 & 0.907 & 0.883 & 1.000 & 1.000 & 1.000 \\ \hline 
	\end{tabular}}
	\caption{Simulated rejection rates in Scenario 5.}
	\label{tablescenario5}
\end{table}
	
	\subsubsection{Analyzing the impact of $b$ and $p$ on MBB and SB}
	
	In order to analyze the influence of parameters $b$ and $p$ on the tests based on MBB and SB, respectively, we run some additional simulations. In particular, we considered Scenario 1 in Section \ref{subsectioned} for two different values of $\delta$, namely $\delta=0$ (null hypothesis) and $\delta=0.075$ (alternative hypothesis). The series length was set to $T=200$. In addition, we fixed different values for both $b$ and $p$. Specifically, we set $b\in\{4, 6, \ldots,20\}$ and  $p\in\{1/4, 1/6, \ldots,1/20\}$. Note that the values employed in Section \ref{subsectioned} for $T=200$ were $b=6$ and $p=1/6$. For each resampling procedure (MBB and SB), dissimilarity measure ($\widehat{d}_{CC}$, $\widehat{d}_B$ and $\widehat{d}_{MLE}$), value of $\delta$ and value of the corresponding input parameter in the selected grid, we repeated the simulation mechanism described in Section \ref{subsectioned} by considering again $N=1000$, $B=500$ and $\alpha=0.05$. 
	
\begin{figure}
		\centering
		\includegraphics[width=1\textwidth]{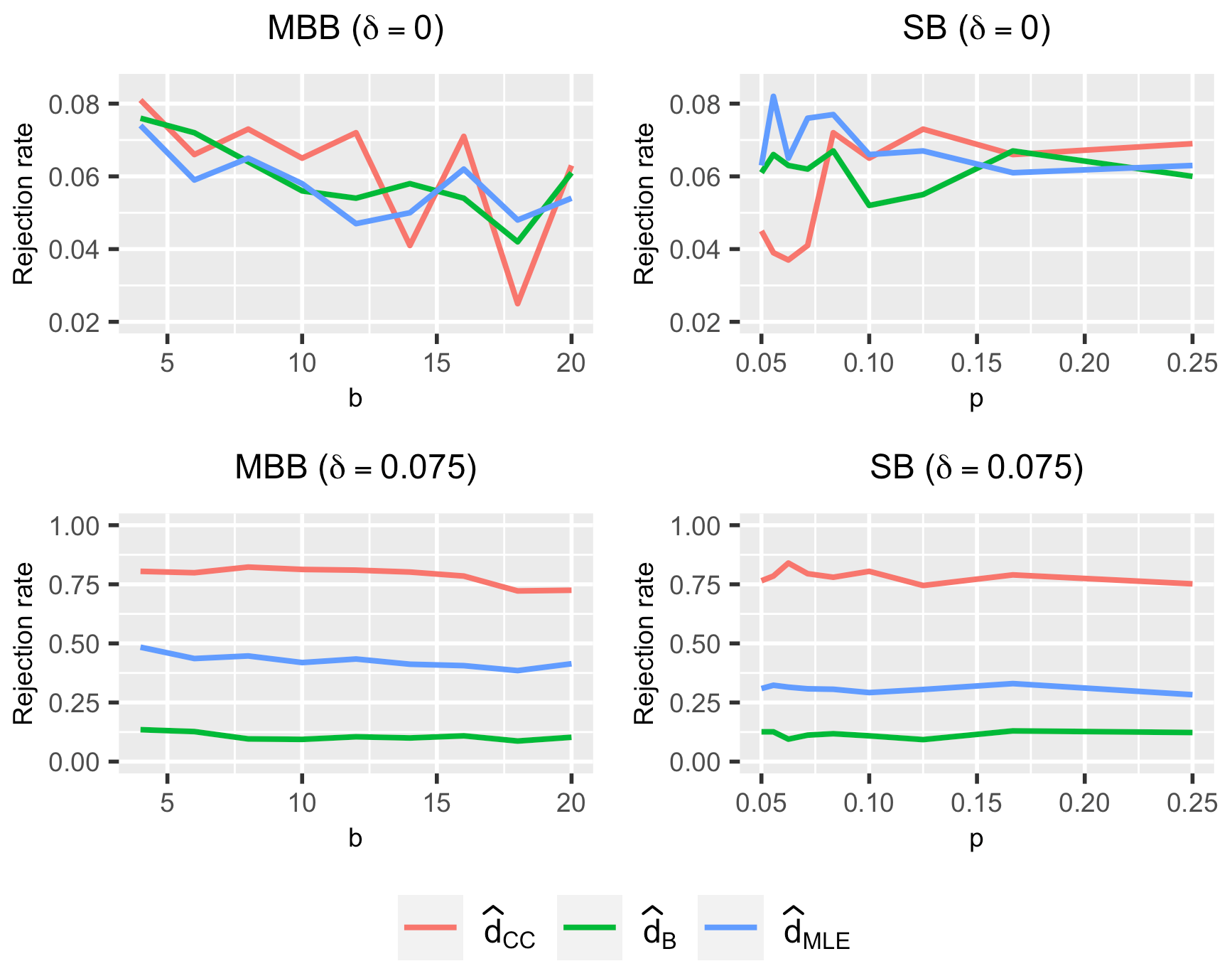}
		\caption{Rejection rates as a function of $b$ and $p$ for procedures MBB (left panels) and SB (right panels). Scenario 1 with $T=200$.}
		\label{mbbsb}
	\end{figure}

	Curves of rejection rates as a function of $b$ (MBB) and $p$ (SB) are displayed in the left and right panels of Figure \ref{mbbsb}, respectively, where each color corresponds to a different dissimilarity measure. In all cases, there are no dramatic differences among the rejection rates associated with different values of the corresponding input parameters. Under the null hypothesis (top panels), the curves oscillate around the nominal level of 0.05 with moderate deviations, which can be due to the noise inherent to the simulation experiments. Analogously, the rejection rates under the alternative hypothesis show a steady behavior for the three metrics and both resampling procedures. Based on previous considerations, one can state that parameters $b$ and $p$ do not have a substantial impact on the behavior of the tests based on MBB and SB when the dependence structures of the underlying processes can be characterized by the first few lags. Note that this is a good property of these procedures, since it frees the user from having to perform hyperparameter selection to obtain suitable values of both parameters, which is usually computationally intensive. 
	
    In sum, the results presented in Section \ref{subsectionfa} corroborate the great performance of the test based on $\widehat{d}_{CC}$ and indicate that a proper selection of parameters $b$ and $p$ is not essential for an appropriate behavior of the resampling procedures MBB and SB.   
	
	\section{Application}\label{sectionapplication}
	
	This section is devoted to show an application of the proposed tests. To that aim, we consider a collection of series that was employed in Section 6.2 of \cite{lopez2023hard} for clustering purposes. Specifically, the dataset contains 40 protein sequences. Proteins are large molecules constituted of one or more chains of simple organic compounds called amino acids. There are 20 different amino acids making up the proteins of any living organism. Therefore, each protein sequence in the considered database can be seen as a CTS with 20 categories. In \cite{lopez2023hard}, the number of categories in each CTS was reduced to 3 by using the so-called  protein sequence encoding. Specifically, the amino acids were categorized into three classes according to its hydrophobicity, which is a common transformation \cite{dubchak1995prediction, dubchak1999recognition}. It is worth highlighting that the application of categorical processes to protein data has been considered in several works \cite{krogh1994hidden, wu2000frequency}. Half of the proteins in the database are found in different parts of human beings, while the other half are present in several variants of COVID-19 virus. The maximum, minimum and median lengths for the CTS in the database are $T=2511$, $T=165$ and $T=426$, respectively.
	
	In \cite{lopez2023hard}, the metrics $\widehat{d}_{CC}$, $\widehat{d}_{B}$ and $\widehat{d}_{MLE}$ were used in combination with the standard partitioning around medoids (PAM) procedure \cite{kaufman2009finding} to perform clustering in the dataset of protein sequences. Specifically, a number of $K=2$ groups was given as input to the PAM algorithm. Thus, note that the main goal of this task was not to obtain groups of series which have been generated from the same stochastic process, but to determine whether the corresponding metrics are able to determine the underlying protein families (human and COVID-19), which are assumed to define the true partition. In order to achieve the former objective, we propose to consider the clustering method based on $p$-values introduced by \cite{maharaj1996significance} along with the hypothesis tests constructed in this manuscript. In particular, the procedure of \cite{maharaj1996significance} is a hierarchical clustering approach starting from a pairwise matrix of $p$-values (which can be seen as a similarity matrix). In fact, a clustering homogeneity criterion for this method is implicitly provided by specifying a threshold significance level $\alpha$ (e.g., 0.05 or 0.01), which automatically determines the number of groups. In this way, those elements with associated $p$-values greater than $\alpha$ will be grouped together, which implies that only those series whose dynamic structures are not significantly different at level $\alpha$ will be placed in the same group. It is worth mentioning that a function implementing this clustering procedure is available by means of the R package \textbf{TSclust} \cite{montero2015tsclust}. 
	
	Based on the above considerations, the clustering method based on the $p$-value previously described was applied to the dataset of protein sequences by considering the 9 hypothesis tests proposed in this paper. For the sake of simplicity and illustration, only the results associated with the metric $\widehat{d}_{CC}$ and the bootstrap approach MBB are provided. Note that the corresponding test showed the best overall performance in the simulation experiments of Section \ref{sectionsimulationstudy}. Computation of the dissimilarity $\widehat{d}_{CC}$ was carried out by considering $\mathcal{L}=\{1, 2\}$, since this set was chosen according to the selection procedure proposed in Section 3.4 of \cite{lopez2023hard}, which is aimed at finding the optimal collection of lags for a joint analysis of a CTS dataset. A straightforward adaptation of the MBB method to the case of series with unequal lengths was considered. A number of $B=500$ bootstrap replicates were used to compute the pairwise matrix of $p$-values and a threshold significance level $\alpha=0.05$ was employed for the hierarchical clustering mechanism. 
	
\begin{figure}
		\centering
		\includegraphics[width=0.8\textwidth]{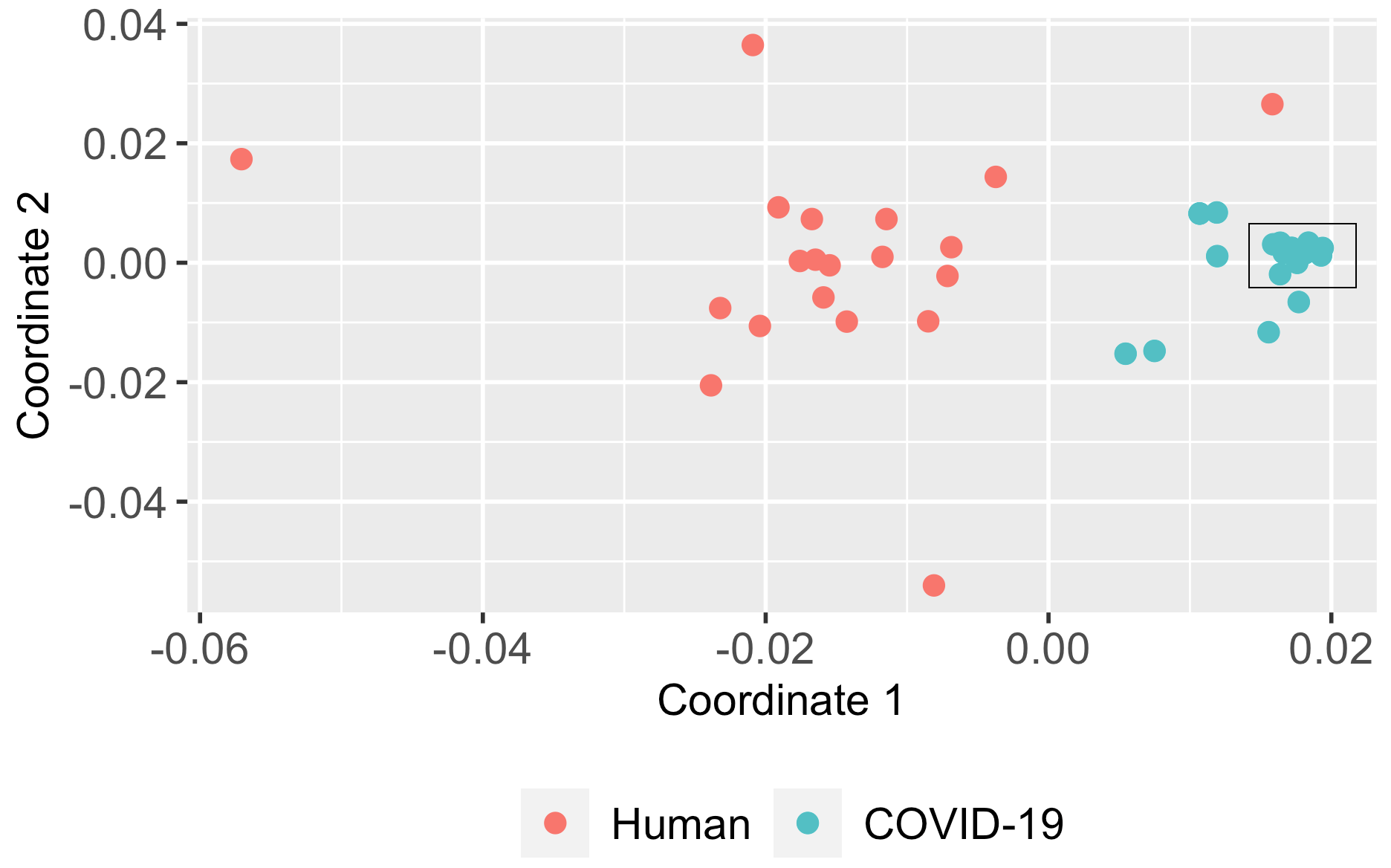}
		\caption{Two-dimensional scaling plane based on distance $\widehat{d}_{CC}$ for the 40 protein sequences. The points inside the rectangle represent time series whose generating processes are not significantly different according to the test based on $\widehat{d}_{CC}$ and MBB.}
		\label{2ds}
	\end{figure}

	As an illustrative step to understand the partition produced by the considered clustering procedure, we performed a two-dimensional scaling (2DS) based on the pairwise dissimilarity matrix for distance $\widehat{d}_{CC}$. In this way, a projection of the protein sequences on a two-dimensional plane preserving the original distances as well as possible is available. The location of the 40 series in the transformed space is displayed in Figure \ref{2ds}. Different colors were used to distinguish human proteins from COVID-19 proteins. 
	
According to the 2DS plot, it is clear that the dissimilarity $\widehat{d}_{CC}$ is able to detect both groups of protein families to some extent. However, these groups exhibit a different degree of variability (e.g., the points representing COVID-19 proteins are more concentrated than the ones representing human proteins). Interestingly, the partition defined by both underlying families is far from being the one identified by the considered clustering approach. In fact, the hierarchical procedure based on $p$-values determines the existence of only one group containing more than one series. Specifically, this group includes thirteen series associated with COVID-19 proteins. These series correspond to the points in Figure \ref{2ds} which lie inside the rectangle. Note that it is reasonable that the generating processes of these time series are not significantly different, since the corresponding pairwise distances are very close to zero in accordance with the 2DS plot. Each one of the remaining series constitutes an isolated group, which indicates rejection of the null hypothesis of equality of generating structures in all their pairwise comparisons. 

It is worth emphasizing that the above application clearly highlights the usefulness of the proposed hypothesis tests. Specifically, we showed that, even in a classical machine learning problem as clustering, an approach based on these tests can lead to dramatically different conclusions than the ones obtained using more conventional techniques. In fact, while a traditional clustering algorithm detects two groups of series displaying similar dependence structures in the protein dataset (those associated with both protein families), the approach based on $p$-values indicates that the series corresponding to human proteins are not so similar, since the equality of generating processes for each pair of them is rejected. Note that the latter approach is more informative and can lead to interesting insights that can not be reached by using standard clustering procedures. 
	
	\section{Conclusions}\label{sectionconclusions}
	
	This work deals with the construction of hypothesis tests for comparing the generating processes of two CTS, which are based on two main elements:
	
	\begin{itemize}
	\item A distance measure between CTS evaluating discrepancy between the marginal distributions and the dependence structures of the series. Specifically, we consider two metrics relying on model-free features ($\widehat{d}_{CC}$ and $\widehat{d}_B$) and a parametric dissimilarity ($\widehat{d}_{MLE}$) assuming a particular class of categorical models. 
	\item A resampling procedure used to properly approximate the asymptotic distribution of the corresponding dissimilarities under the null hypothesis even when this hypothesis is not true. Particularly, we employ a parametric bootstrap approach based on estimated model coefficients and two extensions of the well-known moving blocks bootstrap (MBB) and stationary bootstrap (SB). 
	\end{itemize}

Each combination of dissimilarity measure and resampling procedure gives rise to a different hypothesis test. Both a great ability of the metric to discriminate between underlying structures and a high capability of the resampling mechanism to provide a proper approximation of the corresponding asymptotic distribution are essential to get a good performance of the procedures. The proposed procedures were assessed in a broad simulation study including different types of categorical processes with several levels of complexity. The numerical experiments resulted in the following conclusions:

\begin{itemize}
\item Under the null hypothesis, most tests respect the significance level rather well when a sufficiently large value for the series length is considered. 
\item When the null hypothesis is not true, the test based on $\widehat{d}_{CC}$ and the MBB exhibits the highest power, which is advantageous for the practitioners, since neither the dissimilarity nor the resampling mechanism assume a specific class of categorical model. 
\end{itemize}

The sensibility of methods MBB and SB with respect to their input parameters was also analyzed, and the results indicated that both techniques exhibit approximately the same behavior for a broad range of values for the corresponding parameters. Finally, the test based on $\widehat{d}_{CC}$ and MBB was applied to a dataset of protein sequences along with a clustering procedures based on the $p$-values of the test, and interesting conclusions were reached.  

There are three main ways in which this work can be extended. First, new hypothesis tests similar to the ones proposed here could be constructed by employing additional dissimilarities and resampling procedures. Second, note that bootstrap approaches have to be used in this work due to the impracticality of deriving the asymptotic null distribution of the distances under the general assumption of stationarity. However, by making some additional assumptions (e.g., by considering a specific type of generating models), the computation of the corresponding distributions could be substantially simpler. In such a case, it would be interesting to analyze the advantages and disadvantages of a test based on these distributions with respect to the ones introduced in this manuscript. Third, the clustering methods based on $p$-values applied in Section \ref{sectionapplication} could be rigorously analyzed. In particular, their performance in several simulation scenarios could be assessed by comparing these procedures with alternative clustering approaches.  
	

	\bibliography{mybibfile.bib}
	

\end{document}